\documentclass[%
superscriptaddress,
nofootinbib,
 amsmath,amssymb,
 aps,
 pra,
floatfix,
twocolumn
]{revtex4-2}
\usepackage[utf8]{inputenc}
\usepackage[T1]{fontenc}
\usepackage{babel}
\usepackage{graphicx}
\usepackage{amsfonts}
\usepackage{amssymb}
\usepackage{amsmath}
\usepackage{amsthm}
\usepackage{mathtools}
\usepackage{physics}
\usepackage[normalem]{ulem}
\usepackage{blindtext}

\DeclareMathOperator{\diag}{diag}
\DeclareMathOperator{\const}{const}
\DeclareMathOperator{\re}{re}
\DeclareMathOperator{\im}{im}

\usepackage{braket}
\renewcommand\bra[1]{{\langle{#1}|}}
\renewcommand\ket[1]{{|{#1}\rangle}}
\usepackage{enumerate}
\usepackage{dsfont}
\usepackage{bm}
\usepackage{xcolor}

\newtheorem{theorem}{Theorem}
\newtheorem{proposition}[theorem]{Proposition}

\theoremstyle{definition}
\newtheorem*{example*}{Example}

\usepackage{natbib}
\setcitestyle{square,numbers}
\usepackage[colorlinks]{hyperref}

\begin{document}

\author{Tomasz Linowski}
\affiliation{International Centre for Theory of Quantum Technologies, University of Gdansk, 80-308 Gda{\'n}sk, Poland}
\email[Corresponding author: ]{t.linowski95@gmail.com}

\author{{\L}ukasz Rudnicki}
\affiliation{International Centre for Theory of Quantum Technologies, University of Gdansk, 80-308 Gda{\'n}sk, Poland}
\affiliation{Center for Theoretical Physics, Polish Academy of Sciences, 02-668 Warszawa, Poland}

\title{Classicality of the Bogoliubov transformations and the dynamical Casimir effect through the reduced state of the field}

\date{\today}

\begin{abstract}
We use the reduced state of the field formalism [Entropy \textbf{21}, 705 (2019)] to derive conditions under which a Bogoliubov transformation can be considered semi-classical. We apply this result to the dynamical Casimir effect in a moving medium [Phys. Rev. A \textbf{78}, 042109 (2008)], discussing its classical and quantum features.
\end{abstract}

\maketitle

\section{Introduction}
Arguably one of the most surprising predictions of quantum field theory is the Casimir effect, a physical force arising solely from the presence of quantum fluctuations in the vacuum \cite{Casimir_review_Lamoreaux_2012,Casimir_review_Klimchitskaya_2009,Casimir_review_Kimball_2004}. Since its original formulation in 1948 \cite{Casimir_original_1948}, the phenomenon has garnered a lot of interest, in particular giving rise to many alternative formulations and generalizations. One such generalization, dubbed the dynamical Casimir effect, predicts the spontaneous production of particles in a medium following from non-trivial time dependence of either its boundary or its material coefficients \cite{dynamical_Casimir_Yablonovitch_1989,dynamical_Casimir_Schwinger_1992,dynamical_Casimir_review_Dodonov_2010,dynamical_Casimir_review_Dodonov_2020}. 

In 2008, Iwo Bia{\l}ynicki-Birula\footnote{\textbf{We dedicate this work to Iwo Bia{\l}ynicki-Birula on the occasion of his 90th birthday.}} working together with Zofia Bia{\l}ynicka-Birula\footnote{We also find it a good opportunity to acknowledge the fact that 59 papers from the total of 206 so far published by the Professor, as well as the comprehensive textbook on quantum electrodynamics \cite{QED_Birula}, have been written in this admirable collaboration which started already in 1957 \cite{BullAcad}.} established a third mechanism generating the dynamical Casimir effect: oscillatory motion of a medium \cite{dynamical_Casimir_Birula_2008}. In fact, this mechanism is more general and applies to all kinds of motion, as long as its speed varies in time and one carefully picks the ``incoming'' and ``outgoing'' annihilation and creation operators (see an example of a uniformly accelerated medium \cite{dynamical_Casimir_Rudnicki_2010}). A loosely related phenomenon occurs around large rotating and/or gravitating bodies \cite{IBBgrav}. 

The dynamical Casimir effect is obtained by performing a Bogoliubov transformation: a linear transformation of the creation and annihilation operators of the quantum field preserving canonical commutation relations \cite{qft_2009}. If the Casimir effects are among the most interesting phenomena in quantum theory, Bogoliubov transformations are among its most reliable tools. Originally used to describe superconductivity \cite{Bogoliubov1958,Valatin1958}, today they are widely used in many branches of quantum physics, from optics and theories of magnetism to field theory in a curved spacetime (Unruh effect, Hawking radiation) \cite{optics_bogoliubov,qft_2009,Unruh_effect,Hawking_radiation}.

While the most prominent applications of the Bogoliubov transformations suggest the latter to be inherently quantum, we observe that from the formal point of view, Bogoliubov transformations are essentially equivalent to a change of basis of the Hilbert space. For this reason, one may expect that at least some Bogoliubov transformations could have classical analogues, similarly to local unitary rotations of the Hilbert space, which do not entangle the system. If so, this could shed a new conceptual light on the phenomena described by them.

In this paper we derive an exact set of conditions under which Bogoliubov transformations can be considered semi-classical. By semi-classical (further also referred to as just ``classical'') we understand models which can be described by certain kinetic equations for reduced single-particle states and their displacements --- so called \emph{reduced state of the field} formalism \cite{alicki_reduced_state}. This framework has recently been proved to be an effective tool in probing the classicality of quantum Gaussian evolution \cite{RSF_Linowski_2022}.

In the case of isolated systems, the transformations allowed by our conditions turn out to have a simple interpretation in terms of passive operations, which correspond to classical devices such as beam splitters. In the case of open systems the conditions are less restrictive, which we interpret as some of the total dynamics' ``quantumness'' is being encoded into the environment. Our findings allow us to conduct an in-depth discussion of classicality of the dynamical Casimir effect derived in \cite{dynamical_Casimir_Birula_2008}. We find that, while the overall phenomenon is quantum in nature, the individual photons experience each other as semi-classical dissipative effects. 

This paper is organized as follows. In Section \ref{sec:Casimir_intro}, we introduce the dynamical Casimir effect in moving media. In Section \ref{sec:RSF}, we briefly summarize the most important properties of our main tool -- the reduced state of the field. In Section \ref{sec:Bogoliubov}, we derive our main results: classicality conditions for Bogoliubov transformations. In Section \ref{sec:Casimir_classicality}, we build upon these findings to assess the classicality of the dynamical Casimir effect. We conclude in Section \ref{sec:conclusion}.

\section{Dynamical Casimir effect in a moving medium}
\label{sec:Casimir_intro}
Electromagnetic field is fully described by the set of four three-component vectors: $\vec{D}$ and $\vec{E}$, describing the electric field, along with $\vec{B}$ and $\vec{H}$, describing the magnetic field, which altogether fulfill the Maxwell equations in vacuum \cite{quantum_optics_book_Lambropoulos_2007,gaussian_optics}:
\begin{align} \label{eq:Maxwell_equations}
\begin{split}
    \partial_t{\vec{D}}(\vec{r},t) &= \vec{\nabla} \times {\vec{H}}(\vec{r},t),
        \quad\vec{\nabla}\cdot{\vec{D}}(\vec{r},t) = 0,\\
    -\partial_t{\vec{B}}(\vec{r},t) &= \vec{\nabla} \times {\vec{E}}(\vec{r},t),
        \quad\vec{\nabla}\cdot{\vec{B}}(\vec{r},t) = 0.\\
\end{split}
\end{align}
In the Heisenberg picture, the operators associated to these fields fulfill exactly the same set of equations.

Assuming the field propagates through a homogeneous, isotropic medium moving with a velocity $\vec{v}$ and characterized by constant material coefficients $\mu$, $\epsilon$, the field vectors are related by the Minkowski constitutive relations \cite{Minkowski_1908}
\begin{align} \label{eq:constitutive_relations}
\begin{split}
    {\vec{D}} + \frac{\vec{v}}{c^2} \times {\vec{H}} 
        &= \epsilon\left({\vec{E}}+\vec{v} \times {\vec{B}}\right),\\
    {\vec{B}} - \frac{\vec{v}}{c^2} \times {\vec{E}} 
        &= \mu\left({\vec{H}}-\vec{v} \times {\vec{D}}\right),
\end{split}
\end{align}
where $c$ is the speed of light. 

In the convenient Riemann-Silberstein approach (see a review \cite{Riemann_Silberstein_review_Birula_2013}), the electromagnetic field is combined into two vectors:
\begin{align}
\begin{split}
    {\vec{F}} \coloneqq \frac{1}{\sqrt{2}\epsilon}{\vec{D}}
        + \frac{i}{\sqrt{2}\mu}{\vec{B}},\\
    {\vec{G}} \coloneqq \frac{1}{\sqrt{2}\mu}{\vec{E}}
        + \frac{i}{\sqrt{2}\epsilon}{\vec{H}}.
\end{split}
\end{align}
The advantage of this approach can already be seen in the considered problem, as the constitutive relations (\ref{eq:constitutive_relations}) can be always solved for ${\vec{G}}$, yielding
\begin{align}
\begin{split}
    {\vec{G}} = \frac{c}{n}\left[{\vec{F}}
    +\frac{n^2-1}{c^2 n^2-v^2}\vec{v}\times\left(
    \vec{v}\times{\vec{F}}+icn{\vec{F}}
    \right)\right],
\end{split}
\end{align}
where $n\coloneqq c\sqrt{\epsilon\mu}\geqslant 1$ is the refractive index of the medium. Then, assuming position-independent velocity, $\vec{v}(\vec{r},t)=c\vec{\beta}(t)$, the vacuum Maxwell equations (\ref{eq:Maxwell_equations}) reduce to just one equation:
\begin{align} \label{eq:Maxwell_equation_F}
\begin{split}
    \partial_t{\vec{F}} 
        = &-ic\delta(t)\left[\vec{\beta}(t)\cdot\vec{\nabla}\right]{\vec{F}}
        +\frac{c}{n}\alpha(t)\vec{\nabla}\times{\vec{F}}\\
        &-\frac{c}{n}\delta(t)\vec{\beta}(t)\times\vec{\nabla}
        \left[\vec{\beta}(t)\cdot{\vec{F}}\right],
\end{split}
\end{align}
where
\begin{align}
\begin{split}
    \delta(t) &\coloneqq  \frac{n^2-1}{n^2-\beta^2(t)}, 
        \quad \alpha(t) \coloneqq 1-\delta(t)\beta^2(t).
\end{split}
\end{align}

Under a further assumption that the velocity has a constant direction $\vec{m}$, and with the help of the Fourier decomposition 
\begin{align}
\begin{split}
    \vec{F}(\vec{r},t) = \int &\frac{d^3\vec{k}}{\sqrt{2\pi}^3}
        e^{i\vec{k}\cdot\vec{r}-i\phi(\vec{k},t)} \\ 
        &\times \left[
        \vec{e}(\vec{k})f_+(\vec{k},t)+\vec{e}^*(\vec{k})f_-(\vec{k},t)
        \right],
\end{split}
\end{align}
where $\vec{e}$ are elliptic polarization vectors \cite{dynamical_Casimir_Birula_2008}, the Maxwell equations lead to a pair of ordinary differential equations for the functions $f_\pm$:
\begin{align} \label{eq:f_differential_equations}
\begin{split}
    \partial_t\hat{f}_\pm(\vec{k},t) = \mp i\omega(\vec{k})
        \left[\eta_+(\vec{k},t)\hat{f}_\pm(\vec{k},t)
        -\eta_-(\vec{k},t)\hat{f}_\mp(\vec{k},t)\right],
\end{split}
\end{align}
with
\begin{align}
\begin{split}
    \eta_{\pm}(\vec{k},t) &\coloneqq \frac{1}{2}\left[\frac{\alpha(t)}{\sigma^2(\vec{k})} 
        \pm \sigma^2(\vec{k}) \Delta(\vec{k},t)\right],\\
    \Delta(\vec{k},t) &\coloneqq 1 - \delta(t)\beta^2(t)\cos^2\theta(\vec{k}) ,    
\end{split}
\end{align}
The parameter $\theta$ denotes the angle between the wave vector $\vec{k}$ and the velocity direction $\vec{m}$, while  
$\sigma$ is a \textit{free} real parameter defining the polarization geometry. Last but not least, the phase
\begin{align} \label{eq:phi}
\begin{split}
    \phi(\vec{k},t) &\coloneqq 
        \omega(\vec{k})\cos\theta(\vec{k})\int_{0}^{t} d\tau \delta(\tau)\beta(\tau),
\end{split}
\end{align}
has been extracted to achieve a simplification of the resulting equations.

To obtain the dynamical Casimir effect, it is assumed that the medium is moving with a time-dependent velocity from time $t=0$ up to $t=T$ \cite{dynamical_Casimir_Birula_2008,dynamical_Casimir_Rudnicki_2010}. If the medium just before and after was ``still'' (characterized by $\beta(t)=\mathrm{const}$), the corresponding operators $\hat{f}_\pm$, after a suitable choice of $\sigma$ \cite{dynamical_Casimir_Birula_2008}, can be interpreted in terms of the creation and annihilation operators of photons with right helicity
\begin{align}
\begin{split}
    \hat{f}_+(\vec{k},t)&=\begin{cases}
        \sqrt{\hbar\omega(\vec{k})}\hat{a}_{R,\textnormal{in}}(\vec{k})
            e^{-i\omega(\vec{k})t}, & t<0\\
        \sqrt{\hbar\omega(\vec{k})}\hat{a}_{R,\textnormal{out}}(\vec{k})
            e^{-i\omega(\vec{k})(t-T)}, & t>T        
        \end{cases},
\end{split}
\end{align}
and left helicity
\begin{align}
\begin{split}
    \hat{f}_-(\vec{k},t)&=\begin{cases}
        \sqrt{\hbar\omega(\vec{k})}\hat{a}_{L,\textnormal{in}}^\dag(-\vec{k})
            e^{i\omega(\vec{k})t}, & t<0\\
        \sqrt{\hbar\omega(\vec{k})}\hat{a}_{L,\textnormal{out}}^\dag(-\vec{k})
            e^{i\omega(\vec{k})(t-T)}, & t>T       
        \end{cases}.
\end{split}
\end{align}
Here, $\hat{a}_{L/R,\textnormal{in/out}}$ and their Hermitian conjugates fulfill all the expected properties of the standard annihilation and creation operators. Note that such interpretation is not possible during the acceleration period $t\in[0,T]$ itself, due to impossibility of separation into positive and negative frequency parts.

The final operators are given by the initial ones via the relation \cite{dynamical_Casimir_Birula_2008}:
\begin{align} \label{eq:Casimir_Bogoliubov_transformation}
\begin{split}
    \hat{a}_{R,\textnormal{out}}(\vec{k}) &= e^{-i\phi}
        \left[f_{R+}\hat{a}_{R,\textnormal{in}}(\vec{k})
        +f_{R-}\hat{a}_{L,\textnormal{in}}^\dag(-\vec{k})\right],\\
    \hat{a}_{L,\textnormal{out}}^\dag(-\vec{k}) &= e^{-i\phi}
        \left[f_{L+}\hat{a}_{R,\textnormal{in}}(\vec{k})
        +f_{L-}\hat{a}_{L,\textnormal{in}}^\dag(-\vec{k})\right],\\
\end{split}
\end{align}
where $\phi\equiv \phi(\vec{k},T)$, while $f_{L\pm}\equiv f_{L\pm}(\vec{k},T)$, $f_{R\pm}\equiv f_{R\pm}(\vec{k},T)$ are solutions to the differential equations (\ref{eq:f_differential_equations}) subject to initial conditions
\begin{align} \label{eq:f_initial_conditions}
\begin{split}
    f_{R+}(\vec{k},0) &= f_{L-}(\vec{k},0) = 1, \\
    f_{R-}(\vec{k},0) &= f_{L+}(\vec{k},0) = 0.
\end{split}
\end{align}
It is worth adding that, due to the canonical commutation relations for the outgoing photons (\ref{eq:Casimir_Bogoliubov_transformation})
\begin{align}
\begin{split}
    \big[\hat{a}_{R,\textnormal{out}}(\vec{k}), 
        \hat{a}_{R,\textnormal{out}}^\dag(\vec{k})\big] &= 1,\\
    \big[\hat{a}_{R,\textnormal{out}}(\vec{k}), 
        \hat{a}_{R,\textnormal{out}}(\vec{k})\big] &= 0,
\end{split}
\end{align}
we have
\begin{align} \label{eq:ccr_implies}
\begin{split}
    |f_{R+}|^2 = |f_{R-}|^2 + 1,
\end{split}
\end{align}
with an analogous relation for $f_{L+}$ and $f_{L-}$.

Let us remark that in the original work \cite{dynamical_Casimir_Birula_2008} the functions $f$ were denoted as $f^1_\pm \equiv f_{R\pm}$ and  $f^2_\pm \equiv f_{L\pm}$. Here, we change the notation to make the connection to photon helicity more immediate, as well as to avoid confusing the indeces with exponentiation. We stress, however, that despite corresponding to different photon helicities, the two pairs of functions are interrelated via the initial conditions and have to be considered together.

The Casimir effect is finally obtained by considering system initially in the vacuum and computing the photon number densities after the motion:
\begin{align} \label{eq:photon_densities}
\begin{split}
    \braket{\hat{n}_R(T)} &= \bra{0}
        \hat{a}_{R,\textnormal{out}}^\dag(\vec{k})\hat{a}_{R,\textnormal{out}}(\vec{k})
        \ket{0} = |f_{R-}(\vec{k},T)|^2\delta(0),\\
    \braket{\hat{n}_L(T)} &= \bra{0}
        \hat{a}_{L,\textnormal{out}}^\dag(\vec{k})\hat{a}_{L,\textnormal{out}}(\vec{k})
        \ket{0} = |f_{L+}(\vec{k},T)|^2\delta(0),        
\end{split}
\end{align}
where $\delta(0)$ is the Dirac delta singularity. Note that, due to the symmetry of the evolution equations governing the left and right helicity functions, the two densities are in fact equal:
\begin{align} \label{eq:photon_densities_equal}
\begin{split}
    \braket{\hat{n}_R(T)} = \braket{\hat{n}_L(T)} \equiv \braket{\hat{n}(T)}.        
\end{split}
\end{align}
As was verified in Refs \cite{dynamical_Casimir_Birula_2008,dynamical_Casimir_Rudnicki_2010}, at least for some $\vec{k}$ this number is a growing function of $T$. Therefore, the motion of the medium results in a potentially unbounded particle production in the vacuum and, hence, the prediction of the dynamical Casimir effect.

Transformation (\ref{eq:Casimir_Bogoliubov_transformation}) at the heart of the discussed  phenomenon is an example of a \emph{Bogoliubov transformation} \cite{Bogoliubov1958,Valatin1958}: a linear transformation $\big\{\hat{a}_n,\hat{a}_n^\dag\big\} \to \big\{\hat{a}_n',\hat{a}_n'^\dag\big\}$ of the creation and annihilation operators preserving the canonical commutation relations \cite{qft_2009}. As the main result of this paper, we will derive the precise conditions under which such transformations can be considered semi-classical, with special emphasis put on the classicality of the dynamical Casimir effect in a moving medium.

\section{Reduced state of the field} 
\label{sec:RSF}
To assess the (semi)classicality of Bogoliubov transformations, we first need to define a sensible criterion for what is classical. To this end, we will employ the mesoscopic formalism of the reduced state of the field (RSF) \cite{alicki_reduced_state}, which was already used for similar purposes before \cite{RSF_Linowski_2022}. Since the framework itself is not in the main focus of our study, here, we provide only the basic information about it. For more details, see Ref. \cite{alicki_reduced_state}, where it was introduced by Robert Alicki, Ref. \cite{RSF_Linowski_2022}, where its semi-classical interpretation was established, and Ref. \cite{RSF_Cusumano_2021}, where it was considered in the context of thermodynamics.

We consider an $N$-mode, continuous variable Hilbert space described by set of $N$ annihilation and creation operators $\hat{a}_k$, $\hat{a}_{k'}^\dag$ fulfilling the canonical commutation relations
\begin{align} \label{eq:canonical_commutation_relations}
    \big[\hat{a}_k,\hat{a}_{k'}^\dag\big]=\delta_{kk'},
    \quad \big[\hat{a}_k,\hat{a}_{k'}\big]
    =\big[\hat{a}_k^\dag,\hat{a}_{k'}^\dag\big]=0.
\end{align}
As always, an arbitrary $n$-particle state in the many-body Hilbert space can be constructed by acting on the vacuum state with $n$ appropriate creation operators. Since, in principle, the number of particles in a given mode can be arbitrary, the $N$-mode Hilbert space is infinitely dimensional, and so is the density operator $\hat{\rho}$ constituting the full quantum description of the system.

In some cases, however, the full quantum formalism is not necessary and can be replaced by a simpler, mesoscopic framework. For example, Gaussian states and dynamics can be efficiently studied in the symplectic picture \cite{two-mode_gaussian_etc_proper_norm,using_dissipation_1,using_dissipation_2}. Similarly, to describe macroscopic fields and associated evolution, a formalism called \emph{reduced state of the field (RSF)} has been recently developed \cite{alicki_reduced_state}.

In the RSF framework, instead of the density operator, the system is described by the pair $(r,\ket{\alpha})$. Here, 
\begin{align} \label{eq:single_particle_density_matrix_definition}
    {r}\coloneqq\sum_{k,k'=1}^N
        \Tr\big[\hat{\rho}\,\hat{a}_{k'}^\dag\hat{a}_k\big]
        \ket{k}\bra{k'}
\end{align}
is the \emph{single-particle density matrix}, while the \emph{averaged field} equals
\begin{align} \label{eq:averaged_field_defition}
    \ket{\alpha}\coloneqq\sum_{k=1}^{N}
        \Tr\big[\hat{\rho}\,\hat{a}_k\big]\ket{k}.
\end{align}
The single-particle density matrix contains the simplest non-local information about the system. Additionally, its diagonal elements equal the mean particle numbers: ${r}_{kk} = \braket{\hat{a}_{k}^\dag\hat{a}_{k}}$. Consequently, the matrix is normalized to the mean total particle number. Note that, by construction, the single-particle density matrix is non-negative. The averaged field, on the other hand, contains additional local information.

Much like the previously mentioned symplectic picture requires observables and transformations that are Gaussian, the RSF formalism employs observables that are either \emph{additive} \cite{alicki_reduced_state}:
\begin{align} \label{eq:observable_additive}
    \hat{O} = \sum_{k,k'=1}^N o_{kk'} \hat{a}_k^\dag \hat{a}_{k'}
\end{align}
or \emph{linear}: 
\begin{align} \label{eq:observable_linear}
    \hat{\sigma} = \sum_{k=1}^N \left( \sigma_{k}^* \hat{a}_k + \sigma_k \hat{a}_k^\dag \right).
\end{align}
In the case of macroscopic fields, which are usually modeled as non-interacting fields with dynamics governed by equations linear in creation and annihilation operators, the most relevant observables are of this form. For example, the Hamiltonian is additive, while the position and momentum operators are linear. 

Defining the \emph{reduced observables} corresponding to eqs (\ref{eq:observable_additive}, \ref{eq:observable_linear}) as
\begin{align}
    o = \sum_{k,k'=1}^N o_{kk'} \ket{k} \bra{k'}, 
        \qquad \ket{\sigma} = \sum_{k=1}^N \sigma_{k} \ket{k}
\end{align}
we can indeed see that the associated expectation values can be rewritten in the RSF formalism as \cite{RSF_Linowski_2022}
\begin{align} \label{eq:observables_translation}
\begin{split}
    \Tr \hat{\rho} \, \hat{O} & = \tr r o, \qquad
    \Tr \hat{\rho} \, \hat{\sigma} = \braket{\sigma|\alpha} + \braket{\alpha|\sigma}.
\end{split}
\end{align}

The RSF framework comes equipped with dedicated entropy measures and evolution equations, both derived from the standard quantum description. In the case of entropy, we have the reduced von Neumann and Wehrl entropies \cite{alicki_reduced_state,RSF_Linowski_2022}:
\begin{align} \label{eq:reduced_entropy}
\begin{split}
    s_v(r,\ket{\alpha}) & \coloneqq
        \tr[(r_\alpha+\mathds{1}_N)\ln(r_\alpha+\mathds{1}_N)-r_\alpha\ln r_\alpha],\\
    s_w(r,\ket{\alpha}) & \coloneqq
        \tr\ln(r_\alpha+\mathds{1}_N) + N,        
\end{split}
\end{align}
where $r_\alpha \coloneqq r - \ket{\alpha}\bra{\alpha}$ and $\mathds{1}_N$ denotes the identity matrix in dimension $N$. The reduced entropies arise from applying the maximum entropy principle to the standard von Neumann and Wehrl entropies, respectively \cite{wehrl_entropy_general_properties_Wehrl_1978,wehrl_entropy}.

Finally, RSF evolves according to the \emph{reduced kinetic equations} \cite{alicki_reduced_state,RSF_Linowski_2022}:
\begin{align} \label{eq:evolution_standard_RSF}
\begin{split}
    \frac{d}{dt}{r}=&
        -\frac{i}{\hbar}\big[h,{r}\big]+\ket{\zeta}\bra{\alpha}+\ket{\alpha}\bra{\zeta}\\
    &+\frac{1}{2}\big\{\gamma_\uparrow-\gamma_\downarrow,{r}\big\}+\gamma_\uparrow\\
    &+\sum_j \eta_j\big(u_j{r}u_j^\dag-{r}\big),\\
    \frac{d}{dt}\ket{\alpha}=&
        -\frac{i}{\hbar} h\ket{\alpha}
        +\frac{1}{2}\big(\gamma_\uparrow-\gamma_\downarrow\big)\ket{\alpha}+\ket{\zeta}\\
    &+\sum_j \eta_j\big(u_j-1\big)\ket{\alpha},
\end{split}
\end{align}
which are derived from the Gorini-Kossakowski-Lindblad-Sudarshan (GKLS) equation \cite{GKS_original,lindblad_original} under the assumption that the considered quantum field can be treated as a set of individual particles subject to spontaneous decay and production, as well as interaction with coherent classical sources and random scattering by the environment. The operators entering eq. (\ref{eq:evolution_standard_RSF}) represent:
\begin{itemize}
\item The Hamiltonian:
\begin{align} \label{eq:reduced_h}
    h & \coloneqq \hbar \sum_{k=1}^{N}\omega_k\ket{k}\bra{k}, \quad \omega_k \geqslant 0;
\end{align}
\item Coherent sources:
\begin{align} \label{eq:reduced_zeta}
    \ket{\zeta} & \coloneqq \sum_{k=1}^{N}\zeta_k\ket{k};
\end{align}
\item Particle creation rates:
\begin{align} \label{eq:reduced_gamma}
    \gamma_\uparrow & = \sum_{k,k'=1}^{N}\gamma^{kk'}_{\uparrow}\ket{k}\bra{k'}, 
    \quad\gamma_{\uparrow}\geqslant 0
\end{align}
and analogously particle annihilation rates $\gamma_\downarrow$;
\item Unitary interactions with rates $\eta_j\geqslant 0$, $\sum_j\eta_j=1$:
\begin{align} \label{eq:reduced_u}
    u_j = \sum_{k,k'=1}^{N} u^{kk'}_j\ket{k}\bra{k'},
    \quad u_j^\dag u_j = u_j u_j^\dag = \mathds{1}_N.
\end{align}
For a large number of non-commuting unitaries, the last term in either of the reduced kinetic equations represents random scattering.
\end{itemize}

Note that while not explicitly stated in the original work \cite{alicki_reduced_state}, it is clear from the derivation of the reduced kinetic equations that all the quantities entering it may be time-dependent, provided they fulfill the respective constraints (imposed by complete positivity) during every instant of the evolution.

Although RSF was originally designed to capture the quantum features of macroscopic fields, it has been recently shown to have a convincing interpretation as a semi-classical description of bosonic many-body systems \cite{RSF_Linowski_2022}. For example, it was proved that the RSF formalism contains no information about distillable entanglement in the system, and that both of the reduced entropies are akin to Wehrl's semiclassical entropy \cite{wehrl_entropy}, typically considered as such due to its close association with the phase-space.

Because, by construction, the reduced kinetic equations (\ref{eq:evolution_standard_RSF}) 
preserve the RSF formalism's semi-classical set of degrees of freedom, any time evolution model of the density operator, which can be rewritten as reduced kinetic equations, must be necessarily semi-classical itself. Based on this principle, in Ref. \cite{RSF_Linowski_2022}, quantum Gaussian evolution of light was found to be classical if and only if it consisted strictly of so called passive optical transformations, e.g., beam splitting and phase shifting. Contrary to their remaining active counterparts, such as quantum squeezing, passive transformations can be understood operationally by treating light as a classical wave. In this paper, we adopt a similar methodology for Bogoliubov transformations: if they preserve the set of the degrees of freedom contained within RSF, we will regard them as semi-classical, and if not, we will regard them as inherently quantum.

\section{Classicality of Bogoliubov transformations} 
\label{sec:Bogoliubov}
We are now equipped with the tools necessary to assess the classicality of Bogoliubov transformations. We will consider two distinct cases: Bogoliubov transformations in isolated (closed) systems, and in open systems. The main results of this section are presented in Propositions~\ref{th:Bogoliubov_closed}-\ref{th:Bogoliubov_open}, with proofs in Appendix \ref{app:Bogoliubov}.

\subsection{Isolated system}
In the case of an isolated system, the most general transformation of the density operator is unitary:
\begin{align} \label{eq:Bogoliubov_evolution_total_state}
    \hat{\rho}'=\hat{U}\hat{\rho}\hat{U}^\dag.
\end{align}
For the transformation to be of the Bogoliubov-type, $\hat{U}$ must be such that, for some complex matrix $\mathcal{X}$,
\begin{align} \label{eq:bogoliubov_explicit_general}
    \hat{A}_n'\coloneqq\hat{U}^\dag\hat{A}_n\hat{U} 
        =& \sum_{m=1}^{2N}\mathcal{X}_{nm}\hat{A}_m,
\end{align}
with
\begin{align} \label{eq:generalized_annihilation_operators}
    \hat{A}_n\coloneqq 
    \begin{cases}
        \hat{a}_n, & n\in\{1,\ldots,N\}\\
        \hat{a}_n^\dag, & n\in\{N+1,\ldots,2N\}       
    \end{cases}.
\end{align}
To preserve the canonical commutation relations, the matrix $\mathcal{X}$ has to fulfill the so-called symplectic property \cite{derezinski,napiorkowski}:
\begin{align} \label{eq:symplectic}
    {\mathcal{X}}{\mathcal{S}}{\mathcal{X}}^\dag={\mathcal{S}},
\end{align}
where $\mathcal{S}=\diag\big[\mathds{1}_{N},-\mathds{1}_{N}\big]$. As a consequence of the symplectic property,
\begin{align} \label{eq:X_block_notation}
    \mathcal{X} = \begin{bmatrix}\mathcal{X}_\uparrow&\mathcal{X}_\downarrow
        \\ \mathcal{X}_\downarrow^*&\mathcal{X}_\uparrow^*\end{bmatrix},
\end{align}
where $\mathcal{X}_\updownarrow$ are of size $N\times N$.

Calculating the change in RSF implied by a generic Bogoliubov transformation and forcing the result to be fully contained within the formalism, we obtain the classicality conditions for the closed system Bogoliubov transformations. Furthermore, if the unitary transformation in eq. (\ref{eq:Bogoliubov_evolution_total_state}) depends smoothly on time, then so does the matrix $\mathcal{X}$, turning the discrete Bogoliubov transformation into a continuous \emph{Bogoliubov evolution}. In such case, the density operator can be differentiated with respect to time, and the resulting evolution equation compared with the reduced kinetic equations. 

Proceeding in this way, we obtain our first major result.

\begin{proposition} \label{th:Bogoliubov_closed}
Isolated system Bogoliubov transformations (as described above) are compatible with the RSF formalism and are thus classical with respect to it if and only if
\begin{align} \label{eq:classicality_closed}       
    0=\mathcal{X}_\downarrow.
\end{align}
Additionally, if the transformation depends smoothly on time, the corresponding reduced kinetic equations (\ref{eq:evolution_standard_RSF}) exist and are governed by 
\begin{align} \label{eq:classicality_closed_rke}      
    h = \frac{i\hbar}{2}\left(\frac{d\mathcal{X}_{\uparrow}}{dt}\mathcal{X}_{\uparrow}^{-1}-\mathcal{X}_{\uparrow}^{-\dag}\frac{d\mathcal{X}_{\uparrow}^\dag}{dt}\right).
\end{align}
with the remaining terms vanishing.
\end{proposition}

\begin{proof}
See Appendix \ref{app:Bogoliubov}.
\end{proof}

The obtained classicality condition is easy to interpret: substituting (\ref{eq:classicality_closed}) into the symplectic condition (\ref{eq:symplectic}), we immediately find that $\mathcal{X}$ is also unitary in additional to being symplectic, which means that it is passive. Thus, in a complete analogy to quantum Gaussian evolution \cite{RSF_Linowski_2022}, Bogoliubov transformations in isolated systems are semi-classical only if they correspond to passive transformations.

Let us also remark that, while the absence of the dissipative terms in the obtained reduced kinetic equations was to be expected in an isolated system, the lack of coherent classical sources was not. Indeed, it is easy to see that this lack is not a fundamental property of the Bogoliubov evolution, but rather a consequence of the Bogoliubov transformations (\ref{eq:bogoliubov_explicit_general}) being defined, for simplicity, without constant terms (independent of the creation and annihilation operators).

\subsection{Open system}
In the more general case of an open system, the total density operator of the system and environment (also called bath) is as well transformed according to eq. (\ref{eq:Bogoliubov_evolution_total_state}). However, we are only interested in the state of the system, given by a partial trace over the degrees of freedom of the environment:
\begin{align} \label{eq:Bogoliubov_evolution_system}
    \hat{\rho}_S=\Tr_E \hat{\rho}.
\end{align}
The Bogoliubov transformation itself (\ref{eq:bogoliubov_explicit_general}) remains the same. Still, assuming the system and the environment span $N_S$ and $N_E$ modes respectively, it is convenient to additionally split the matrices entering the block decomposition (\ref{eq:X_block_notation}) into
\begin{align} \label{eq:X_arrows_block_notation}
    \mathcal{X}_\uparrow = \begin{bmatrix}
        \mathcal{X}_{\uparrow S} & \mathcal{X}_{\uparrow C} \\
        \mathcal{X}_{\uparrow C'} & \mathcal{X}_{\uparrow E}
        \end{bmatrix},
    \quad \mathcal{X}_\downarrow = 
        \begin{bmatrix}
        \mathcal{X}_{\downarrow S} & \mathcal{X}_{\downarrow C} \\ 
        \mathcal{X}_{\downarrow C'} & \mathcal{X}_{\downarrow E}
        \end{bmatrix},
\end{align}
where $\mathcal{X}_{\updownarrow S}$ is an $N_S\times N_S$ matrix associated with the system, $\mathcal{X}_{\updownarrow E}$ is an $N_E\times N_E$ matrix associated with the environment, and $\mathcal{X}_{\updownarrow C}$, $\mathcal{X}_{\updownarrow C'}$ are appropriately-sized matrices associated with both. Note that the case of the closed system can be retrieved easily by setting $N_E=0$ (which, in particular, implies $\mathcal{X}_{\updownarrow}=\mathcal{X}_{\updownarrow S}$) and dropping the then-redundant lower indices $S$.

For a generic initial state of the bath-system ensemble, the dynamics of the latter cannot be separated from the dynamics of the former, making it impossible to even compare with the RSF formalism. Nonetheless, even in this completely general setting, we were able to derive necessary conditions for classicality of Bogoliubov transformations.

\begin{proposition} \label{th:Bogoliubov_open_necessary}
Open system Bogoliubov transformations (as described above) can be compatible with the RSF formalism and thus be classical with respect to it only if
\begin{align} \label{eq:classicality_open}
    0=\mathcal{X}_{\downarrow S}.
\end{align}
\end{proposition}

\begin{proof}
See Appendix \ref{app:Bogoliubov}.
\end{proof}

Unlike the condition (\ref{eq:classicality_closed}) for the closed system, the classicality condition for the open system is difficult to interpret. However, comparing it with its closed system counterpart, we can at least see that the latter is much more restrictive: it requires the whole matrix $\mathcal{X}_\downarrow$ to vanish, while the former requires only its system part $\mathcal{X}_{\downarrow S}$ to vanish. Therefore, depending on how we define the degrees of freedom of the system, we may find the same total dynamics to be either classical or quantum from the point of view of the system. This will indeed be the case in the next section, where we will find that the dynamical Casimir effect falls exactly into this category.

Still, any such interpretation has to be made with care, since it must be stressed that the condition (\ref{eq:classicality_open}) is not equivalent to classicality, but only necessary for it. In stark contrast to the closed system, in the case of open system, whether or not a given Bogoliubov transformation is classical from the point of view of RSF depends not only on the matrix $\mathcal{X}$ defining it, but also on the total initial state of the system-environment ensemble. It is possible that, for particularly strongly correlated total initial states, the only semi-classical Bogoliubov transformations are those that induce completely separate dynamics for the system and environment, essentially defying the notion of an open system.

To make stronger statements, we are therefore forced to make some restrictions. Firstly, we assume that the initial total state is separable with respect to the bipartition between the system and the bath. This is a typical assumption in the theory of quantum open systems. In particular, the GKLS equation cannot be derived without it \cite{quantum_open_systems_separability}. Since, in particular, the reduced kinetic equations governing the time evolution in the RSF formalism are derived from a GKLS equation, it is only natural to also make this assumption in the present case. 

Secondly, we assume that the bath is initially in the vacuum state. Note that, while this assumption is a very strong one, it is fulfilled by many well-studied and useful models, such as quantum limited amplification, quantum limited attenuation and phase conjugation channels, utilized, e.g. in studies of Gaussianity, entropy and entanglement \cite{example_transformations,De_Palma_2017,squashed_entanglement_de_Palma_2019}. More importantly for us, as we will discuss in the next section, it is also satisfied by the dynamical Casimir effect.

Under the above assumptions, we obtain our final main result for Bogoliubov transformations.

\begin{proposition} \label{th:Bogoliubov_open}
The classicality condition (\ref{eq:classicality_open}) is both necessary and sufficient for open system Bogoliubov transformations with the environment initially in the vaccum state. Additionally, if such transformations depend smoothly on time, the corresponding reduced kinetic equations exist provided
\begin{align} \label{eq:Bogoliubov_open_gammas_condition}
    \mathcal{W}\geqslant 0, 
        \quad \mathcal{W}-\mathcal{Y}_r \geqslant 0
\end{align}
and are governed by 
\begin{align} \label{eq:Bogoliubov_open_RSF_correspondence}
    h=-\hbar\mathcal{Y}_{i}/2, 
        \quad \gamma_\downarrow=\mathcal{W}, 
        \quad \gamma_\uparrow=\mathcal{W}-\mathcal{Y}_r,
\end{align}
with the remaining terms vanishing. Here,
\begin{align} \label{eq:Bogoliubov_open_Y,W_definitions}
\begin{split}
    \mathcal{Y}_{i}&\coloneqq-i
        \left(\mathcal{Y}-\mathcal{Y}^\dag\right), \quad
    \mathcal{Y}\coloneqq
        \frac{d\mathcal{X}_{\uparrow S}}{dt}\mathcal{X}_{\uparrow S}^{-1}, \\
    \mathcal{Y}_{r}&\coloneqq
        \mathcal{Y}+\mathcal{Y}^\dag, \quad\qquad\:
    \mathcal{D}\coloneqq \mathcal{X}_{\downarrow C}\mathcal{X}_{\downarrow C}^\dag, \\
    \mathcal{W}&\coloneqq
        \frac{d\mathcal{D}}{dt}-\mathcal{Y}\mathcal{D}-\mathcal{D}\mathcal{Y}^\dag.
\end{split}
\end{align}
\end{proposition}

\begin{proof}
See Appendix \ref{app:Bogoliubov}.
\end{proof}

Interestingly, the obtained Bogoliubov reduced kinetic equations do not depend on any components of the matrix $\mathcal{X}$ labeled by the subscripts $C'$, despite depending on the components labeled by $C$. At first, this may appear surprising, since a priori both are equally responsible for describing the correlations between the system and the environment. The asymmetry is resolved by interpreting the $C$ components as encoding the influence of the environment on the system, and the $C'$ components as encoding the influence of the system on the environment. The lack of the $C'$ components in the description of the system then becomes expected. As an additional argument for this view, we observe that if we exchanged the roles of the system and the environment, the equations would depend on the $C'$ components, with the $C$ components missing.

Proposition \ref{th:Bogoliubov_open} will be our main tool in the study of classicality of the dynamical Casimir effect. Before we do it, however, let us illustrate our results so far with a short but instructive example: the Gaussian amplification process.

\begin{example*}[Gaussian amplification process] 
\label{ex:amplification} 
In the Gaussian amplification process, an arbitrary initial state of the $N$-mode system
\begin{align}
    \hat{\rho}(t_0)=\int \frac{d^{2N}\vec{z}_0}{\pi^N} P_0(\vec{z}_0)
        \ket{\vec{z}_0}\bra{\vec{z}_0}
\end{align}
is driven by a heat bath into the state \cite{amplification}
\begin{align}
\begin{split}
    \hat{\rho}(t)&=\int \frac{d^{2N}\vec{z}_0}{\pi^N} P_0(\vec{z}_0)
        \bigotimes_{j=1}^N\int
        \frac{d^{2}z_j}{\pi}\rho_j(t)\ket{z_j}\bra{z_j},\\
    \rho_j(t)&\coloneqq \frac{1}{n_{j}(t)} e^{-|z_j-z_{0j}e^{\kappa_j t}|^2/n_{j}(t)}. 
\end{split}
\end{align}
Here, the integration is over the real and imaginary parts of the complex vectors $\vec{z}_0$, $\vec{z}$; $P_0(\vec{z}_0)$ denotes the Glauber–Sudarshan P representation \cite{P_representation_Glauber,P_representation_Sudarshan} of the initial state, $\ket{z_j}$ are coherent states, $\kappa_j$ is the amplification rate of the $j$-th mode and 
\begin{align}
    n_{j}(t)\coloneqq(1+m_j)\left(e^{2\kappa_j t}-1\right),
\end{align}
where $m_j$ is the mean number of photons in the $j$-th mode of the bath, assumed to be effectively constant throughout the whole process (this is true as long as the bath is much bigger than the system).

The corresponding RSF can be easily calculated:
\begin{align}
\begin{split}
    {r}_{kk'}(t)&=\int \frac{d^{2N}\vec{z}_0}{\pi^N} P_0(\vec{z}_0)
        \prod_{j=1}^N\int \frac{d^{2}z_j}{\pi}\rho_j(t)
        z_{k}z_{k'}^*,\\
    \alpha_k(t)&=\int \frac{d^{2N}\vec{z}_0}{\pi^N} P_0(\vec{z}_0)
        \prod_{j=1}^N\int \frac{d^{2}z_j}{\pi}\rho_j(t)
        z_{k}.
\end{split}
\end{align}
The integrals over $z_j$ can be performed using the standard result \cite{altland}:
\begin{align} \label{eq:integration_trick}
    \int \frac{d^{2N}\vec{z}}{\pi^N}e^{-\vec{z}^\dag{\mu}\vec{z}
        +\vec{s}^\dag\vec{z}+\vec{z}^\dag\vec{s}}
        =\frac{1}{\det\mu}e^{\vec{s}^\dag{\mu}^{-1}\vec{s}},
\end{align}
where $\mu$ denotes an invertible matrix and $\vec{s}$ is a vector of size $N$. In our case
\begin{align}
\begin{split}
    \mu^{-1} &= n(t) \coloneqq  \sum_{j=1}^N n_j(t)\ket{j}\bra{j},\\
    \vec{s} &= n^{-1}(t)\ket{\vec{z}_0(t)},
    \quad \ket{\vec{z}_0(t)}\coloneqq\sum_{j=1}^N z_{0j}e^{\kappa_j t}\ket{j}.
\end{split}
\end{align}
This yields
\begin{align} \label{eq:evol_amplification}
\begin{split}
    {r}(t)&=n(t)
        +\braket{\ket{\vec{z}_0(t)}\bra{\vec{z}_0(t)}}_0,\\
    \ket{\alpha(t)}&=\braket{\ket{\vec{z}_0(t)}}_0,
\end{split}
\end{align}
where $\braket{\cdot}_0 \coloneqq \int \frac{d^{2N}\vec{z}_0}{\pi^N} P_0(\vec{z}_0)(\cdot)$. The formulae (\ref{eq:evol_amplification}) induce the following differential evolution equations:
\begin{align} \label{eq:evol_in_amplification_final}
\begin{split}
    \frac{d}{dt}{r}&=\frac{1}{2}
        \big\{2{\kappa}\left({\mathds{1}}+m\right)-2{\kappa}m,{r}\big\}
        +2{\kappa}\left(\mathds{1}+m\right),\\
    \quad\frac{d}{dt}\ket{\alpha}
        &=\frac{1}{2}\big(2{\kappa}\left({\mathds{1}}+m\right)-2{\kappa}m\big)\ket{\alpha},
\end{split}
\end{align}
where $m\coloneqq\sum_{j=1}^N m_j\ket{j}\bra{j}$ and $\kappa\coloneqq\sum_{j=1}^N\kappa_j\ket{j}\bra{j}$. Clearly, the equations have the form of reduced kinetic equations (\ref{eq:evolution_standard_RSF}) with ${\gamma}_\uparrow=2{\kappa}\left({\mathds{1}}+m\right)$, ${\gamma}_\downarrow=2{\kappa}m$ and $h=\ket{\zeta}=\mu(du)=0$.

According to Proposition \ref{th:Bogoliubov_open_necessary}, any open system Bogoliubov evolution that can be represented by reduced kinetic equations has to necessarily fulfill the classicality condition (\ref{eq:classicality_open}). To see that this is indeed the case in the Gaussian amplification process, we observe that it is generated by a Bogoliubov transformation of the form \cite{example_transformations}
\begin{align}
\begin{split}
    \mathcal{X}_{\uparrow}&=\cosh\kappa t
    \begin{bmatrix}
    \mathds{1}_N & 0 \\
    0 & \mathds{1}_N
    \end{bmatrix},\quad
    \mathcal{X}_{\downarrow}=\sinh\kappa t
    \begin{bmatrix}
    0 & \mathds{1}_N \\
    \mathds{1}_N & 0
    \end{bmatrix},
\end{split}
\end{align}
Clearly, $\mathcal{X}_{\downarrow S}$, being the upper left-hand side block component of $\mathcal{X}_{\downarrow}$, vanishes, as required by the aforementioned condition.

The fact that we found the Gaussian amplification process to be semi-classical is not surprising: intuitively, Gaussian amplification can be interpreted as pumping particles into the system, until it reaches essentially macroscopic size. The process is well known for turning quantum phenomena into more classical ones. For example, it was previously shown that the Glauber-Sudarshan P distribution of an infinitely amplified state approaches the semi-classical Husimi Q distribution \cite{quantum_phase_Q_amplification_Schleich_1992,Q_representation}. Similarly, the von Neumann entropy of the maximally amplified state approaches the semi-classical Wehrl entropy \cite{De_Palma_2017,wehrl_entropy_general_properties_Wehrl_1978}. More recently, it has been shown that the amplified Pegg-Barnett phase formalism approaches the Paul phase formalism \cite{Pegg-Barnett_Paul_relation_Linowski}.
\end{example*}

\section{Classicality of the dynamical Casimir effect} 
\label{sec:Casimir_classicality}
Armed with the classicality conditions (\ref{eq:classicality_closed}, \ref{eq:classicality_open}), we are now ready to come back to the dynamical Casimir effect. We begin by observing that, while the phenomenon spans an infinite number of modes of photons with both helicities, its defining Bogoliubov transformation (\ref{eq:Casimir_Bogoliubov_transformation}) couples them in pairs only. Any mode $\vec{k}$ of the right helicity photons is coupled only to itself and the mode $-\vec{k}$ of the left helicity photons. For this reason, we can restrict our analysis to two modes, with no loss in generality.

Written in terms of the matrix $\mathcal{X}$, the Bogoliubov transformation (\ref{eq:Casimir_Bogoliubov_transformation}) reads
\begin{align} \label{eq:X_Casimir}
    \mathcal{X} = \begin{bmatrix}
        e^{-i\phi}f_{R+} & 0 & 0 & e^{-i\phi}f_{R-} \\
        0 & e^{i\phi}f_{L-}^* & e^{i\phi}f_{L+}^* & 0 \\
        0 & e^{i\phi}f_{R-}^* & e^{i\phi}f_{R+}^* & 0 \\
        e^{-i\phi}f_{L+} & 0 & 0 & e^{-i\phi}f_{L-}
    \end{bmatrix}.
\end{align}
The classicality interpretation depends on what we consider to be the system. 

In the most natural view, the system spans photons with both left and right helicity. Hence, we have a closed, two-mode system. Comparing eq. (\ref{eq:X_Casimir}) with (\ref{eq:X_block_notation}), we easily find the classicality criterion (\ref{eq:classicality_closed}) to read explicitly
\begin{align} \label{eq:classicality_Casimir_closed}
     f_{R-}(\vec{k},T) = 0 = f_{L+}(\vec{k},T).
\end{align}
Looking at eq. (\ref{eq:photon_densities}), we can immediately see that this implies no Casimir effect, i.e. the photon production in the vacuum is zero. Thus, according to the RSF formalism, any dynamical Casimir effect is necessarily non-classical, as expected.

To see the physical reason for this, we go back to the differential equations (\ref{eq:f_differential_equations}), along with the initial conditions (\ref{eq:f_initial_conditions}). It is easy to see that eq. (\ref{eq:classicality_Casimir_closed}) can be fulfilled if and only if $\eta_-(\vec{k},t)=0$. This is equivalent to $\sigma(\vec{k})=[\alpha/\Delta(\vec{k})]^{1/4}$, where, due to the time-independence of $\sigma$, $\alpha$ and $\Delta$ have to be time-independent too, implying constant velocity. The equations for the remaining functions can be then easily solved, yielding \cite{dynamical_Casimir_Birula_2008}
\begin{align}
    f_{R+}(\vec{k},t) = f_{L-}^*(\vec{k},t) = e^{-i\tilde{\omega}(\vec{k})t},
\end{align}
where $\tilde{\omega}=\omega\sqrt{\alpha\Delta}$. Substituting this into eq. (\ref{eq:Casimir_Bogoliubov_transformation}), we find that the final creation and annihilation operators simplify to just
\begin{align}
\begin{split}
    \hat{a}_{R,\textnormal{out}}(\vec{k}) &= 
        e^{-i\left[\phi(\vec{k},T)+\tilde{\omega}(\vec{k})T\right]}
        \hat{a}_{R,\textnormal{in}}(\vec{k}),\\
    \hat{a}_{L,\textnormal{out}}^\dag(-\vec{k}) &= 
        e^{-i\left[\phi(\vec{k},T)-\tilde{\omega}(\vec{k})T\right]}
        \hat{a}_{L,\textnormal{in}}^\dag(-\vec{k}),\\
\end{split}
\end{align}
i.e. they are multiplied by a phase. Obviously, this phase is irrelevant for the expectation values of the corresponding number operators on the vacuum, which is why the dynamical Casimir effect cannot take place for constant velocities.

However, there is another point of view. Nothing stops us from interpreting exclusively the left helicity photons as the system, and the right helicity photons as the environment. Then, we are dealing with an open one-mode system subject to influence from a one-mode environment. By comparing (\ref{eq:X_Casimir}) with eqs (\ref{eq:X_block_notation}, \ref{eq:X_arrows_block_notation}), we immediately find that now, the classicality condition (\ref{eq:classicality_open}) always holds, regardless of the form of the functions $f_{R\pm}$, $f_{L\pm}$. Crucially, because the mode associated with the right helicity photons is initially in the vacuum state, then, due to Proposition \ref{th:Bogoliubov_open}, this classicality condition is both necessary and sufficient. Does this mean that the Casimir effect is, in the end, classical? Or maybe it means that the RSF formalism is not a valid tool for probing classicality after all?

In our opinion, neither. Consider, for example, the maximally entangled two-qubit Bell state \cite{bell_original,quantum_information_book}:
\begin{align} \label{eq:two_qubit}
    \ket{\Phi_+} \coloneqq \frac{1}{\sqrt{2}}\left(\ket{00}+\ket{11}\right).
\end{align}
If, in an analogy to the Casimir effect, we consider only the first qubit as the system, we will find it to be in the maximally mixed state:
\begin{align}
    \hat{\rho}_S = \Tr_{2\textnormal{nd qubit}} \ket{\Phi_+}\bra{\Phi_+} 
        = \frac{1}{2}\hat{\mathds{1}}_2,
\end{align}
which can certainly be considered classical. Of course, this does not mean that the Bell state that we started with was classical. Instead, its ``quantumness'' was contained in the correlations between the two qubits, rather than any of the two qubits themselves.

In the case of the Casimir effect, and the Bogoliubov transformations in general, it is even more apparent what happens with the quantumness. Consider the matrix element $\mathcal{X}_{\downarrow 12} = \mathcal{X}_{\downarrow C} = f_{R-}(\vec{k},T)$, which in our case encodes the correlations between photons with left and right helicities. For a generic initial state, these correlations are potentially quantum. Thus, if a closed system is to be considered classical, they must necessarily vanish: $\mathcal{X}_{\downarrow 12} = \mathcal{X}_{\downarrow C}=0$, as they constitute an integral part of the system. However, in the case of an open system, the discussed correlations are no longer part of the system, and instead enter it only at the level of the environmental effects, most easily seen through the evolution eq. (\ref{eq:Bogoliubov_open_Y,W_definitions}). Therefore, even if they have a strictly quantum origin, the system experiences them only as dissipation, which in this case happens to have a semi-classical interpretation in terms of particle annihilation and creation rates.

Alternatively, we can think of the Casimir process as consisting of two parts. The first, captured by the matrix $\mathcal{X}_\uparrow$, describes the morphing of photons with left helicity into those with right helicity, and vice versa. The second, captured by the matrix $\mathcal{X}_\downarrow$, describes the creation of photons with both helicities. The former, being semi-classical, is unconstrained by the RSF formalism. The latter, however, being more quantum in nature, is forbidden by RSF, unless the quantumness can be encoded into the environment, as discussed previously.

Finally, let us observe that even though the Bogoliubov transformation (\ref{eq:Casimir_Bogoliubov_transformation}) is technically of the discrete type, as the creation and annihilation operators are formally ill-defined during the acceleration period $t\in[0,T]$, the functions $f_{R\pm}$, $f_{L\pm}$ defining the transformation are well defined at all times. Adding to that the fact that the final moment of acceleration $T$ is completely arbitrary, we can consider eq. (\ref{eq:Casimir_Bogoliubov_transformation}) as defining a smooth Bogoliubov evolution in the parameter $T$.

Since, as explained previously, the initial total state fulfills the requirements of Proposition \ref{th:Bogoliubov_open}, the Bogoliubov evolution at hand must have a representation in terms of the reduced kinetic equations (\ref{eq:evolution_standard_RSF}) with eq. (\ref{eq:Bogoliubov_open_RSF_correspondence}) at the input. Indeed, making use of the latter equation, we find
\begin{align} \label{eq:Casimir_evolution}
\begin{split}
    h & = \hbar\omega
        \left(\eta_+ + \eta_- \re\frac{f_{R-}}{f_{R+}} +\delta\beta\cos\theta\right),\\
    \gamma_\uparrow &= 2\omega\eta_-
        \frac{|f_{R-}|^2}{|f_{R+}|^2} \im\frac{f_{R+}}{f_{R-}},\\
    \gamma_\downarrow &= 0.
\end{split}
\end{align}
For more details regarding the derivation of these identities, see Appendix \ref{app:Casimir_evolution}. Here, we focus on their physical significance. 

To start with, we note that, as expected, the Hamiltonian for the photons is proportional to their frequency. Furthermore, the particle annihilation rate is zero, which intuitively corresponds to the fact that the dynamical Casimir effect results only in the spontaneous creation of particles, not their disappearance. Finally, once again abusing the differential equations (\ref{eq:Bogoliubov_open_RSF_correspondence}), we can easily calculate that the time derivative of the total photon density (\ref{eq:photon_densities_equal}) equals 
\begin{align}
\begin{split}        
    \frac{d}{dT}\braket{\hat{n}} = 2\omega \eta_- |f_{R-}|^2 \im\frac{f_{R+}}{f_{R-}},
\end{split}
\end{align}
which, using eqs (\ref{eq:ccr_implies}, \ref{eq:Casimir_evolution}), can be rewritten as simply
\begin{align} \label{eq:photon_number_final_result}
\begin{split}        
    \frac{d}{dT}\braket{\hat{n}} = \gamma_\uparrow \left(\braket{\hat{n}}+1\right).
\end{split}
\end{align}
This result has three worthwhile implications. 

Firstly, it has a sound physical interpretation: the time derivative of the total photon density in the dynamical Casimir effect turns out to be simply proportional to the current photon density times the current particle creation rate. Secondly, it tells us that the non-negativity of $\gamma_\uparrow$, which is required for the result to be physical, is equivalent to the non-negativity of photon number growth. In particular, because of the initial condition (\ref{eq:f_initial_conditions}), a valid matrix $\gamma_\uparrow$ by its very construction prevents negative photon numbers. Finally, because of the $\braket{\hat{n}}$-independent term on the r.h.s., our final result (\ref{eq:photon_number_final_result}) proves that the dynamical Casimir effect occurs for any non-zero $\gamma_\uparrow$, which can be traced to any non-constant velocity of the medium ($\gamma_\uparrow=0$ holds only for $\eta_-=0$, which holds only for $\vec{\beta} = \const$).

\section{Concluding remarks} 
\label{sec:conclusion}
In this paper, we employed the recent mesoscopic formalism of the reduced state of the field to derive the exact conditions under which Bogoliubov transformations in either isolated or open systems should be considered semi-classical. Applying our result to the case of dynamical Casimir effect in the medium moving with a varying speed, we found that, while the photons with left and right helicity see each other as semi-classical objects, the Casimir effect itself is genuinely quantum, as expected. Let us stress that the analysis is made possible because for each wave vector we can consider two polarization degrees of freedom. Therefore, it is essential that the described phenomenon is ``\emph{based on full Maxwell equations in three dimensions}'' as pointed out at the end of the Conclusions section in Ref. \cite{dynamical_Casimir_Birula_2008}.

\begin{acknowledgements}
We acknowledge support by the Foundation for Polish Science (International Research Agenda Programme project, International Centre for Theory of Quantum Technologies, Grant No. 2018/MAB/5, cofinanced by the European Union within the Smart Growth Operational program).
\end{acknowledgements}

\bibliography{report}
\bibliographystyle{obib}

\appendix

\section{Proofs of Propositions \ref{th:Bogoliubov_closed}-\ref{th:Bogoliubov_open}} \label{app:Bogoliubov}
\setcounter{equation}{0}
\renewcommand{\theequation}{\ref{app:Bogoliubov}\arabic{equation}}
In this appendix, we prove our main results regarding the classicality of Bogoliubov transformations: Propositions \ref{th:Bogoliubov_closed}-\ref{th:Bogoliubov_open}.

To this end, in addition to RSF, we will employ two auxiliary mesoscopic fields. The first, defined originally in \cite{RSF_Linowski_2022}, is the \emph{conjugate RSF}:
\begin{align} \label{eq:conjugate_RSF}
\begin{split}
    {c}&\coloneqq\sum_{k,k'=1}^{N}
        \Tr\big[\hat{\rho}\,\hat{a}_{k'}\hat{a}_k\big]
        \ket{k}\bra{k'},\\
    \ket{\alpha^*}&\coloneqq\sum_{k=1}^{N}
        \Tr\big[\hat{\rho}\,\hat{a}_k^\dag\big]
        \ket{k}.
\end{split}
\end{align}
The second is the \emph{generalized RSF}:
\begin{align} \label{eq:generalized_RSF}
\begin{split}
    {g}&\coloneqq\sum_{k,k'=1}^{2N}
        \Tr\big[\hat{\rho}\,\hat{A}_{k'}^\dag\hat{A}_k\big]
        \ket{k}\bra{k'},\\
    \ket{\mathcal{A}}&\coloneqq\sum_{k=1}^{2N}
        \Tr\big[\hat{\rho}\,\hat{A}_k\big]
        \ket{k}.
\end{split}
\end{align}
It is easy to see that the three reduced fields are related to each other as follows
\begin{align} \label{eq:reduced_fields_relations}
    {g}=\begin{bmatrix} {r} & {c} \\ {c}^* & {r}^T+\mathds{1}_N\end{bmatrix},
        \quad \ket{\mathcal{A}} = \ket{\alpha}\oplus\ket{\alpha^*}.
\end{align}
We add that, by definition, ${r}={r}^\dag$, ${c}={c}^T$ and $\ket{\alpha}^*=\ket{\alpha^*}$.

\subsection{Proof of Proposition \ref{th:Bogoliubov_closed}}
We start with Proposition \ref{th:Bogoliubov_closed}. It is easy to see that, due to eqs (\ref{eq:generalized_RSF}, \ref{eq:Bogoliubov_evolution_total_state}, \ref{eq:bogoliubov_explicit_general}), under a generic Bogoliubov transformation, the generalized RSF $({g},\ket{\mathcal{A}})$ transforms as
\begin{align} \label{eq:Bogoliubov_evolution_initial_state_dependency}
    {g}'=\mathcal{X}{g}\mathcal{X}^\dag,
        \qquad\ket{\mathcal{A}'}=\mathcal{X}\ket{\mathcal{A}}.
\end{align}
Eqs (\ref{eq:reduced_fields_relations}, \ref{eq:X_block_notation}) then imply
\begin{align} \label{eq:Bogoliubov_open_evolution_RSF_quantum}
\begin{split}
    {r}'=\:&\mathcal{X}_{\uparrow}{r}\mathcal{X}_{\uparrow}^\dag
        +\mathcal{X}_{\uparrow}{c}\mathcal{X}_{\downarrow}^\dag
        +\mathcal{X}_{\downarrow}{c}^\dag\mathcal{X}_{\uparrow}^\dag
        +\mathcal{X}_{\downarrow}\big[{r}^T_S+\mathds{1}_{N}\big]
            \mathcal{X}_{\downarrow}^\dag,\\
    \ket{\alpha'}=\:&\mathcal{X}_{\uparrow}\ket{\alpha}
        +\mathcal{X}_{\downarrow}\ket{\alpha^*},
\end{split}
\end{align}
Clearly, this couples RSF to the conjugate field, meaning that it does not preserve the set of the associated degrees of freedom. For an arbitrary initial state the coupling vanishes only if eq. (\ref{eq:classicality_closed}) is fulfilled, which is what we wanted to show.

Assuming the time-dependent case with the classicality condition (\ref{eq:classicality_closed}) fulfilled, eq. (\ref{eq:Bogoliubov_open_evolution_RSF_quantum}) reduces to 
\begin{align} \label{eq:Bogoliubov_open_evolution_RSF}
\begin{split}
    {r}(t)=\mathcal{X}_{\uparrow}(t){r}(t_0)\mathcal{X}_{\uparrow}^\dag(t),\quad
    \ket{\alpha(t)}=\mathcal{X}_{\uparrow}(t)\ket{\alpha(t_0)}.
\end{split}
\end{align}
These equations are reversible:
\begin{align} \label{eq:Bogoliubov_open_evolution_RSF_reversed}
\begin{split}
    {r}(t_0)=\mathcal{X}_{\uparrow}^{-1}(t){r}(t)\mathcal{X}_{\uparrow}^{-\dag}(t),\quad
    \ket{\alpha(t_0)}=\mathcal{X}_{\uparrow}^{-1}(t)\ket{\alpha(t)}.
\end{split}
\end{align}
Taking the time derivative of eq. (\ref{eq:Bogoliubov_open_evolution_RSF}) and making use of eq. (\ref{eq:Bogoliubov_open_evolution_RSF_reversed}) we obtain the reduced kinetic equations (\ref{eq:evolution_standard_RSF}) with eq. (\ref{eq:classicality_closed_rke}) at the input. This concludes the proof.

\subsection{Proof of Proposition \ref{th:Bogoliubov_open_necessary}}
To prove Proposition \ref{th:Bogoliubov_open_necessary}, we observe that the reduced fields of the total state of the system and the environment have the structure
\begin{align} \label{eq:reduced_fields_system_environment_relation}
\begin{split}
    {r}&=
    \begin{bmatrix}
        {r}_S & {r}_{C} \\
        {r}_{C}^\dag & {r}_{E}
    \end{bmatrix},\quad
    {c}=
    \begin{bmatrix}
        {c}_S & {c}_{C} \\
        {c}_{C}^T & {c}_{E}
    \end{bmatrix},\\
    \ket{\alpha}&=\ket{\alpha^*}^*=\ket{\alpha_S}\oplus\ket{\alpha_E},
\end{split}
\end{align}
where $(r_S,\ket{\alpha_S})$, $(c_S,\ket{\alpha_S^*})$ are the reduced fields of the system, $(r_E,\ket{\alpha_E})$, $(c_E,\ket{\alpha_E^*})$ are the reduced fields of the environment and ${r}_{C}$, ${c}_{C}$ contain the system-bath correlations. This fact follows directly from the definitions of the fields. For example,
\begin{align}
\begin{split}
    ({r}_S)_{kk'}&\coloneqq\Tr\big[\big(\Tr_E\hat{\rho}\big)
        \hat{a}_{k'}^\dag\hat{a}_{k}\big]
        =\Tr\left[\hat{\rho}\,\hat{a}_{k'}^\dag\hat{a}_{k}\right]\eqqcolon {r}_{kk'}.
\end{split}
\end{align}
The remaining relations are proved in a similar fashion. 

For a generic initial total state, the dynamics are quite complex. Making use of the block-form decompositions (\ref{eq:reduced_fields_system_environment_relation}, \ref{eq:X_arrows_block_notation}) in eq. (\ref{eq:Bogoliubov_open_evolution_RSF}), we obtain the a rather lengthy expression for the transformed RSF of the system, which can be written as
\begin{align} \label{eq:open_RSF_transformation_general}
\begin{split}
    {r}_S'=\:& F_{\uparrow\uparrow}(r) + F_{\downarrow\uparrow}(c^*)
        + F_{\uparrow\downarrow}(c) + F_{\downarrow\downarrow}(r^T+1),\\
    \ket{\alpha_S'}=\:&\mathcal{X}_{\uparrow S}\ket{\alpha_S}
        +\mathcal{X}_{\uparrow C}\ket{\alpha_E}
        +\mathcal{X}_{\downarrow S}\ket{\alpha^*_S}
        +\mathcal{X}_{\downarrow C}\ket{\alpha^*_E},
\end{split}
\end{align}
where
\begin{align}
\begin{split}
    F_{ab}(x) \coloneqq\:& \mathcal{X}_{a S} x_S \mathcal{X}_{b S}^\dag
        + \mathcal{X}_{a S} x_C \mathcal{X}_{b C}^\dag   \\
        &+ \mathcal{X}_{a C} x_C^\dag \mathcal{X}_{b S}^\dag
        + \mathcal{X}_{a C} x_E \mathcal{X}_{b C}^\dag.
\end{split}
\end{align}
Similarly to case with the closed system transformation, eq. (\ref{eq:open_RSF_transformation_general}) may preserve the set of the degrees of freedom associated with the RSF formalism in the system only if it does not depend on the conjugate field of the system, $(c_S,\ket{\alpha^*_S})$. Close inspection of eq. (\ref{eq:open_RSF_transformation_general}) reveals that this is possible only if eq. (\ref{eq:classicality_open}) is fulfilled, which is what we wanted to prove.

Let us stress, however, that this condition is merely necessary for the RSF degrees of freedom to be preserved. Depending on the state of the bath, the remaining fields $r_c$, $c_C$, $r_E$, $c_E$ will in general cause the system to go beyond the RSF framework. In the most radical case, the equations may preserve the formalism's set of the degrees of freedom of only if all terms dependent on these additional fields vanish, reducing the system-environment ensemble to two separate closed systems.

\subsection{Proof of Proposition \ref{th:Bogoliubov_open}}
Finally, to prove Proposition \ref{th:Bogoliubov_open}, we note that, as is easy to calculate from their definitions, the initial reduced fields with the environment initially in the vacuum state fulfill
\begin{align}
\begin{split}
    r_C = r_E = c_C = c_E = 0, \quad \ket{\alpha_E} = \ket{\alpha^*_E} = 0.
\end{split}
\end{align}
Plugging this into eq. (\ref{eq:open_RSF_transformation_general}), we find that it simplifies to
\begin{align} \label{eq:r_Bogoliubov_vacuum}
\begin{split}
    {r}_S'&=\mathcal{X}_{\uparrow S}{r}_S\mathcal{X}_{\uparrow S}^\dag
        +\mathcal{X}_{\downarrow C}\mathcal{X}_{\downarrow C}^\dag, \\
    \ket{\alpha_S'}&=\mathcal{X}_{\uparrow S}\ket{\alpha_S},
\end{split}
\end{align}
where we assumed the classicality condition (\ref{eq:classicality_open}). Clearly, the final field depends only on the initial RSF, preserving the associated degrees of freedom. Therefore, in this case the condition (\ref{eq:classicality_open}) is not only necessary, but also sufficient for classicality.

It remains to show that if the transformation depends smoothly on time, the corresponding reduced kinetic equations are given by eq. (\ref{eq:Bogoliubov_open_RSF_correspondence}). In the time-dependent case, eq. (\ref{eq:r_Bogoliubov_vacuum}) becomes
\begin{align} \label{eq:r_Bogoliubov_vacuum_t}
\begin{split}
    {r}_S(t)&=\mathcal{X}_{\uparrow S}(t){r}_S(0)\mathcal{X}_{\uparrow S}^\dag(t)
        +\mathcal{X}_{\downarrow C}(t)\mathcal{X}_{\downarrow C}^\dag(t), \\
    \ket{\alpha_S(t)}&=\mathcal{X}_{\uparrow S}(t)\ket{\alpha_S(0)}.
\end{split}
\end{align}
These relations are reversible:
\begin{align} \label{eq:r_Bogoliubov_t_0}
\begin{split}
    {r}_S(0)&=\mathcal{X}_{\uparrow S}^{-1}(t)\left[{r}_S(t)
        -\mathcal{X}_{\downarrow C}(t)\mathcal{X}_{\downarrow C}^\dag(t)\right]
        \mathcal{X}_{\uparrow S}^{-\dag}(t), \\
    \ket{\alpha_S(0)}&=\mathcal{X}_{\uparrow S}^{-1}(t)\ket{\alpha_S(t)}.
\end{split}
\end{align}
Differentiating eq. (\ref{eq:r_Bogoliubov_vacuum_t}) with respect to time, making use of eq. (\ref{eq:r_Bogoliubov_t_0}) and rearranging the terms we arrive at the differential evolution equations:
\begin{align} \label{eq:Bogoliubov_open_RSF_final}
\begin{split}
    \frac{d}{dt}{r}&=\frac{1}{2}[\mathcal{Y}_{i},{r}]
        +\frac{1}{2}\big\{\mathcal{Y}_{r},{r}\big\}+\mathcal{W},\\
    \frac{d}{dt}\ket{\alpha}&=\frac{1}{2}\mathcal{Y}_{i}\ket{\alpha}
        +\frac{1}{2}\mathcal{Y}_{r}
        \ket{\alpha},
\end{split}
\end{align}
where the matrices $\mathcal{Y}_{r}$, $\mathcal{Y}_{i}$, $\mathcal{W}$ are as defined in eq. (\ref{eq:Bogoliubov_open_Y,W_definitions}). Clearly, the derived equations have the form of the reduced kinetic equations characterized by eq. (\ref{eq:Bogoliubov_open_RSF_correspondence}). Thus, they describe valid dynamics provided the $\gamma_{\updownarrow}$ matrices are non-negative, as required by eq. (\ref{eq:Bogoliubov_open_gammas_condition}). This concludes the proof.

\section{Proof of eq. (\ref{eq:Casimir_evolution})} \label{app:Casimir_evolution}
\setcounter{equation}{0}
\renewcommand{\theequation}{\ref{app:Casimir_evolution}\arabic{equation}}
In this appendix, we derive the explicit forms of the operators (\ref{eq:Casimir_evolution}) governing the reduced kinetic equations for the dynamical Casimir effect.

By comparing eq. (\ref{eq:X_Casimir}) with eqs (\ref{eq:X_block_notation}, \ref{eq:X_arrows_block_notation}), we immediately identify
\begin{align}
    \mathcal{X}_{\uparrow S} = e^{-i\phi}f_{R+}, 
        \quad \mathcal{X}_{\downarrow C} = e^{-i\phi}f_{R-}.
\end{align}
Plugging this into eq. (\ref{eq:Bogoliubov_open_Y,W_definitions}) and then eq. (\ref{eq:Bogoliubov_open_RSF_correspondence}), on the way utilizing the differential equations (\ref{eq:f_differential_equations}), we obtain, after a lengthy but straightforward calculation,
\begin{align} \label{eq:Casimir_evolution_prototype}
\begin{split}
    h & = \hbar\omega
        \left(\eta_+ + \eta_- \re\frac{f_{R-}}{f_{R+}}\right) + \hbar\frac{d\phi}{dt},\\
    \gamma_\uparrow &= 2 \omega\eta_- |f_{R-}|^2
        \left(\im\frac{f_{R+}}{f_{R-}} + \im\frac{f_{R-}}{f_{R+}}\right),\\
    \gamma_\downarrow &= 2 \omega\eta_- 
        \left[|f_{R-}|^2\im\frac{f_{R+}}{f_{R-}} + 
        \left(|f_{R-}|^2+1\right)\im\frac{f_{R-}}{f_{R+}}
        \right].
\end{split}
\end{align}
It remains to show that these formulas reduce to eq. (\ref{eq:Casimir_evolution}).

In the case of the Hamiltonian, all we need to do is to differentiate eq. (\ref{eq:phi}) with respect to time. Due to the Leibniz integral rule,
\begin{align}
\begin{split}
    \frac{d\phi(\vec{k},t)}{dt} =
        \omega(\vec{k})\delta(t)\beta(t)\cos\theta(\vec{k}),
\end{split}
\end{align}
from which we immediately see that the first lines of eqs (\ref{eq:Casimir_evolution_prototype}, \ref{eq:Casimir_evolution}) coincide.

As for $\gamma_\uparrow$, we observe that for any complex number $w$
\begin{align} \label{eq:complex_trivial}
\begin{split}
    \im (w^{-1}) = -\frac{\im w}{|w|^2}.
\end{split}
\end{align}
Taking $w = f_{R+}/f_{R-}$, we get
\begin{align} \label{eq:gamma_up_almost}
\begin{split}
    \gamma_\uparrow &= 2 \omega\eta_- |f_{R-}|^2
        \left(1 - \frac{|f_{R-}|^2}{|f_{R+}|^2}\right) \im \frac{f_{R+}}{f_{R-}}.
\end{split}
\end{align}
Using eq. (\ref{eq:ccr_implies}) and simplifying, we quickly find that the second lines of eqs (\ref{eq:Casimir_evolution_prototype}, \ref{eq:Casimir_evolution}) also coincide.

Finally, we have to show that $\gamma_\downarrow = 0$. Once again utilizing the relation (\ref{eq:ccr_implies}), we obtain
\begin{align}
\begin{split}
    \gamma_\downarrow = 2 \omega\eta_- 
        |f_{R-}|^2\left[\im\frac{f_{R+}}{f_{R-}} + 
        \frac{|f_{R+}|^2}{|f_{R-}|^2}\im\frac{f_{R-}}{f_{R+}}
        \right].
\end{split}
\end{align}
It is easy to see that the bracketed term vanishes upon the use of eq. (\ref{eq:complex_trivial}).

\end{document}